\documentclass[a4paper,11pt]{article}
\pdfoutput=1 

\usepackage{jcappub} 

\usepackage[T1]{fontenc} 

\usepackage{amsmath}
\usepackage{graphicx}
\usepackage{subcaption}
\usepackage{soul}
\usepackage[colorinlistoftodos]{todonotes}
\usepackage{multirow}

\title{Requirements for future CMB satellite missions: photometric and band-pass response calibration}

\author[a,b,1]{T. Ghigna,\note{Corresponding author.}}
\author[b]{T. Matsumura,}
\author[c]{G. Patanchon,}
\author[d]{H. Ishino,}
\author[e,f,b,g]{M. Hazumi}

\affiliation[a]{Sub-Department of Astrophysics, University of Oxford, Keble Rd, Oxford OX1 3RH, UK}
\affiliation[b]{Kavli Institute for the Physics and Mathematics of the Universe (Kavli IPMU, WPI), UTIAS, The University of Tokyo, Kashiwa, Chiba 277-8583, Japan}
\affiliation[c]{Université de Paris, CNRS, Astroparticule et Cosmologie, F-75013 Paris, France}
\affiliation[d]{Department of Physics, Okayama University, 3-1-1 Tsushimanaka, Kita-ku, Okayama 700-8530, Japan}
\affiliation[e]{High Energy Accelerator Research Organization (KEK), Tsukuba, Ibaraki 305-0801, Japan}
\affiliation[f]{Institute of Space and Astronautical Science (ISAS), Japan Aerospace Exploration Agency (JAXA), Sagamihara, Kanagawa 252-0222, Japan}
\affiliation[g]{The Graduate University for Advanced Studies (SOKENDAI), Miura District, Kanagawa 240-0115, Hayama, Japan}

\emailAdd{tommaso.ghigna@physics.ox.ac.uk}

\abstract{Current and future Cosmic Microwave Background (CMB) Radiation experiments are targeting the polarized $B$-mode signal. The small amplitude of this signal makes a successful measurement challenging for current technologies. Therefore, very accurate studies to mitigate and control possible systematic effects are vital to achieve a successful observation. An additional challenge is coming from the presence of polarized Galactic foreground signals that contaminate the CMB signal. When they are combined, the foreground signals dominate the polarized CMB signal at almost every relevant frequency. Future experiments, like the LiteBIRD space-borne mission, aim at measuring the CMB $B$-mode signal with high accuracy to measure the tensor-to-scalar ratio $r$ at the $10^{-3}$ level.
We present a method to study the photometric calibration requirement needed to minimize the leakage of polarized Galactic foreground signals into CMB polarization maps for a multi-frequency CMB experiment. We applied this method to the LiteBIRD case, and we found precision requirements for the photometric calibration in the range $\sim10^{-4}-2.5\times10^{-3}$ depending on the frequency band. Under the assumption that the detectors are uncorrelated, we found requirements per detector in the range $\sim0.18\times10^{-2}-2.0\times10^{-2}$. Finally, we relate the calibration requirements to the band-pass resolution to define constraints for a few representative band-pass responses: $\Delta\nu\sim0.2-2$ GHz.}

\begin{document}
\maketitle
\flushbottom

\section{Introduction}
\label{sec:intro}

The Cosmic Microwave Background (CMB) Radiation has played a primary role for cosmology in the past 50 years \cite{penzias_wilson}. 
It has been a powerful tool to test, confirm or rule out cosmological models and our understanding of the fundamental physical laws of the universe we live in. 
Even after more than 50 years of intensive studies, more accurate observations can provide better measurements of the cosmological parameters and help constrain inflation, improving our knowledge of the origin and composition of our universe.

In the past two decades, most of the efforts of the CMB international community have converged on the study of the CMB polarized signal. 
This weak signal is a tracer of the physics governing the primordial plasma, which our universe was made of soon after the Big Bang, up until the Recombination era. 
There are two processes that can generate polarization in the CMB: scalar modes and tensor modes. The latter are equivalent to a background of primordial gravitational waves generated during inflation. It is common to quantify the amplitude of tensor modes in terms of the tensor-to-scalar ratio parameter $r$. A precise measurement of $r$ will allow us to shed new light on the physics of the early Universe and possibly constrain inflation.

The current main target of CMB experiments is to measure primordial $B$-modes \cite{SimonsObs, bicep1, CMBs4SinceBook, litebird1, litebird2, LiteBIRD_Sugai}, which are generated at the Last Scattering Surface through the interaction of photons with a background of gravitational waves potentially arising through inflation \cite{kamionkowski, Seljak_1997a, Seljak_1997b, Zaldarriaga1998}. $B$-modes can also be generated after Last Scattering because of the conversion of $E$-modes through weak gravitational lensing. Because of the angular scale of the large scale structure, this effect becomes mainly relevant on small angular scales.

A number of ground based experiments, including but not limited to POLARBEAR \cite{polarbear1}, ACT \cite{ACT1} and SPT \cite{SPT1}, have already successfully measured the $B$-mode spectrum on small angular scales. 
However, only a large angular scale measurement can probe directly the inflationary paradigm, by measuring both the reionization bump ($\ell \lesssim 10$), where the primordial $B$-mode signal is expected to be larger than the lensing signal for $r \gtrsim 10^{-3}$, and the recombination bump ($10 \lesssim \ell \lesssim 200$).

The BICEP/Keck experiment \cite{bicep1} has been operating from the South Pole for the past decade, and has been able to set an upper limit of $r<0.06$ with a 95\% confidence level by combining its data with those from Planck and WMAP \cite{bicep2018}. The lowest $\ell$-range ($\ell \lesssim 20$) remains anyway inaccessible from the ground, therefore the only way to observe both the predicted reionization ($\ell \lesssim 10$) and recombination bumps of the $B$-mode spectrum, with an accuracy on $r\sim10^{-3}$, is a space-based observation, thanks to the availability of the whole sky and the absence of the atmosphere.

Moreover, recent studies using data from the Planck satellite \cite{PlanckEarly, Planck2013Legacy, Planck2015Legacy, planck2018Legacy} have pointed out that contamination from Galactic polarized emissions is not negligible anywhere on the sky, even away from the Galactic plane. Therefore a careful multi-frequency analysis of the whole sky is the only way to give a final and unbiased answer to the search for primordial $B$-modes \cite{Krachmalnicoff2016}. A second important lesson learned from the experience of the Planck experiment is the relevance of systematic effects for CMB polarization measurements. The Planck team identified several systematic effects. For the sake of this paper, we mention the band-pass mismatch effect described in \cite{PlanckHFI2015}, which has been the main driver for the study presented in this paper. Band-pass mismatch can cause total intensity to polarization leakage as described in \cite{Hoang_2017, Banerji_2019}, and polarized foreground leakage through component separation, which is the focus of this paper.

The LiteBIRD mission \cite{litebird1, litebird2, LiteBIRD_Sugai} is under design with the goal of measuring the primordial $B$-mode signal with a sensitivity (in terms of tensor-to-scalar ratio) $\sigma_{r} \leq 0.001$ (including systematics and statistical uncertainties), with 3 years of observation from the second Sun-Earth Lagrangian point. Observations will cover a wide frequency range from 34 to 448 GHz, divided into 15 frequency bands. 
The broad frequency coverage is justified by the requirement of measuring the Galactic foregrounds in order to be able to characterize them with high accuracy and separate them from the underlying cosmological signal.
Although useful for foreground separation, having many frequency bands requires accurate inter-frequency calibration. 
An imperfect photometric calibration or poor band-pass knowledge may cause leakage of foreground signals into the estimated CMB maps. These effects have been the subject of other studies like the one in \cite{Ward_2018}, although with a slightly different approach and extension compared to what we report in this paper.

In order to set photometric calibration and band-pass resolution requirements, we have performed simulations of component separation in the map (pixel) domain. Propagating the effect of an imperfect calibration (photometric or band-pass resolution) to the maps, we estimate the impact on the reconstruction of the tensor-to-scalar ratio. We can then set requirements to minimize the bias on the recovered cosmological parameter. 
With this procedure, we can test the combined effect of instrumental systematics and contamination from Galactic sources (synchrotron and dust) at different frequencies, to find which bands are more sensitive to miscalibration and hence define the calibration requirements.
Assuming the CMB dipole as the main photometric calibrator, we can also define the band-pass resolution necessary to minimize the effect of an imperfect color correction due to the presence of the Galactic foregrounds. This study is particularly important because it can guide the experiment to select the most appropriate observation strategy to suppress photometric calibration uncertainty, and drive the design of the ground calibration system (most likely a Fourier Transform Spectrometer) used to characterize the band-pass response of the telescope in order to achieve the required resolution.

In Section \ref{sec:formalism} we describe the instrumental model for a CMB polarimeter, the formalism used throughout the paper, the sky model assumed for the analysis, and the method to propagate the calibration uncertainty to the frequency maps. 
In Section \ref{sec:systematics} we guide the reader through the analysis method applied to estimate the impact on the tensor-to-scalar ratio reconstruction, and in Section \ref{sec:results} we show the results of the analysis. Finally, in Section \ref{sec:discussion} we discuss the results and their implications. 

\section{Formalism} \label{sec:formalism}
In this section we describe the analytical formalism used in this study: the instrumental model, the sky modeling assumptions and how we define the effect of band-pass response on the data.
\subsection{Instrumental model}
\label{sec:observations}
For the scope of this paper, we can assume all telescope parameters to be ideal, static, frequency-independent and perfectly known, with the exception of the band-pass response. We need to stress here that this is an approximation in order to focus on the photometric and band-pass calibration accuracy. For example, we do not take into account a frequency-dependent telescope beam. With this approximation we can write the instrumental model for a single detector on the focal plane of a CMB polarimeter as:
\begin{equation}\label{eq:bandpassOnly}
    d = \int d\nu~G(\nu) \Big\{I(\nu,\hat{n}) + \Big[Q(\nu,\hat{n})\cos2\psi(\nu)+U(\nu,\hat{n})\sin2\psi(\nu)\Big]\Big\} + n\footnote{Since we assume a perfectly known beam function $B(\nu, \Omega)$, we omit the antenna effective area $A_e=\lambda^2/\Omega_b$ ($\Omega_b=\int d\Omega B(\nu, \Omega)$), and the integration over the solid angle $\int d\Omega$.},
\end{equation}
where $G(\nu)$ is the band-pass response, $I(\nu,\hat{n})$, $Q(\nu,\hat{n})$ and $U(\nu,\hat{n})$ represent the total intensity and polarized intensity of the sky, $n$ is the detector noise and $\psi$ represents the orientation of the polarization sensitive detector on the sky. 
When the polarimeter employs a polarization modulator such as a rotating Half-Wave Plate (HWP), we can rewrite $\psi(\nu)$ in Equation \ref{eq:bandpassOnly} as $2\rho-\psi(\nu)$ where $\rho$ is the HWP angle \cite{Matsumura2009}. In this paper, the difference between the two expressions is not relevant for the scope of the discussion. 

For completeness we highlight here a few aspects concerning the band-pass response. Experiments like Planck reconstruct the polarization pattern (Q and U) by differencing orthogonal detector pairs. This approach is simple but susceptible to mismatches such as differences in the band-pass responses; in which case the different band-pass response between the 2 orthogonal detectors leads to total intensity leakage into the final Q and U maps \cite{Hoang_2017, Banerji_2019}. This effect can be mitigated using a polarization modulator (for example a rotating Half-Wave Plate) as the first optical element of the telescope (as in the ABS experiment \cite{KusakaABS1,KusakaABS2}). In this configuration, it is possible to demodulate the signal of a single detector to measure both the Q and U parameters simultaneously. This second approach is immune to mismatch between orthogonal detectors, and therefore, the main contribution to the uncertainty in the data comes from the finite knowledge of the band-pass response, as it will be clear in the following sections. 

In this paper we do not take into account I to P leakage (total intensity to polarization) because we assume here an experiment using an ideal polarization modulator. See \cite{Hoang_2017, Banerji_2019}, for a discussion of the effect of I to P leakage in the absence of a polarization modulator.

\subsection{Sky  model}\label{skyModel}
The sky emission at frequency $\nu$ and position $\hat{n}$, can be modelled as a sum of the CMB signal and Galactic foregrounds \cite{Hoang_2017}:
\begin{eqnarray}
I(\nu,\hat{n}) &=& I_0(\nu)+\left.\frac{\partial B(\nu,T)}{\partial T}\right|_{T_0} \Delta T_{cmb}(\hat{n}) + \sum_f I_f(\nu,\hat{n}) \label{eq:skyI}\\
Q(\nu,\hat{n}) &=& \left.\frac{\partial B(\nu,T)}{\partial T}\right|_{T_0} \Delta Q_{cmb}(\hat{n}) + \sum_f Q_f(\nu,\hat{n}) \label{eq:skyQ}\\
U(\nu,\hat{n}) &=& \left.\frac{\partial B(\nu,T)}{\partial T}\right|_{T_0} \Delta U_{cmb}(\hat{n}) + \sum_f U_f(\nu,\hat{n}) \label{eq:skyU}
\end{eqnarray}
where $I_0$ represents the CMB monopole with temperature $T_0=2.7255$ K, and the second term on the right hand side of Equation \ref{eq:skyI} is the anisotropy of the CMB: $B(\nu,T)$\footnote{In equation \ref{eq:bandpassOnly} we have already implicitly multiplied by the antenna effective area $A_b=\lambda^2/\Omega_b$ and computed the integral over the solid angle and the, therefore in this scenario the black-body function is in units of W/Hz.} is the black-body spectrum and $\Delta T_{cmb}(\hat{n})$ is the temperature fluctuation around $T_0$. Finally, the last term is the combination of every other relevant sky component $f$ (i.e. thermal dust, synchrotron, etc). In the same way, in Equations \ref{eq:skyQ} and \ref{eq:skyU} we define the two Stokes parameters for polarization without the monopole term.

We take into account two Galactic foreground emissions: thermal dust and synchrotron. We model the thermal dust emission (in spectral radiance units) with a modified black-body (or grey-body) with spatially-uniform spectral index $\beta_d=1.55$, for total intensity, and $1.6$ for polarization and temperature $T_d=19.6$ K \cite{Planck2018IV, Planck2018XI, Hensley_2018}:
\begin{equation}
\label{eq:dust}
    [I_{d}, Q_{d}, U_{d}](\nu, \hat{n})=A_{[I,Q,U],d}(\nu_0, \hat{n})\Big(\frac{\nu}{\nu_0}\Big)^{\beta_d}\frac{B(\nu,T_d)}{B(\nu_0,T_d)}
\end{equation}
and synchrotron (in spectral radiance units) with a power law with uniform spectral index $\beta_s=-1.1$ and without curvature:
\begin{equation}
\label{eq:synch}
    [I_{s}, Q_{s}, U_{s}](\nu, \hat{n})=A_{[I,Q,U],s}(\nu_0, \hat{n})\Big(\frac{\nu}{\nu_0}\Big)^{\beta_s}.
\end{equation}
In Equations \ref{eq:dust} and \ref{eq:synch} $\nu_0$ is a pivot frequency necessary to define the amplitude of the foreground signals.

In this study we do not take into account contributions from free-free emission, spinning dust and carbon monoxide (CO) transition lines, because previous experiments have shown negligible polarization levels ($\lesssim 1\%$, see \cite{DICKINSON20181, ichiki2014, Greaves_1999, Puglisi_2017}). We also do not account for point sources because they can be masked during the analysis.

\subsection{Systematic effect due the band-pass uncertainty}\label{sec:framework}


In \cite{Hoang_2017, Banerji_2019} the effect of total intensity to polarization leakage (I to P) due to both the scanning strategy and band-pass uncertainty has been already discussed extensively, especially in the case of a polarimeter that does not employ a rotating HWP. The authors showed that this effect becomes negligible when using a polarization modulator, thanks to the absence of band-pass mismatch between orthogonal detectors and a more uniform scanning-angle ($\psi$ in Equation \ref{eq:bandpassOnly}) distribution that helps reducing the deviation of the cross-link matrix from its ideal case (see \cite{Hoang_2017} for details).

In this paper we discuss another effect arising from a limited knowledge of the band-pass response, we call it polarized foreground leakage, in analogy with total intensity to polarization leakage. This effect has the same impact for both polarimeter architectures, with or without a polarization modulator. As explained in the following, the band-pass knowledge is closely related to the photometric calibration accuracy of the data, therefore we developed a framework to include both effects.

A poor photometric calibration accuracy of the data can cause an imperfect estimation of the foreground components signal, and therefore a leakage of these into the recovered CMB map. Since polarized dust and synchrotron are brighter than the polarised CMB $B$-mode signal at all frequencies, this leakage might impact dramatically the ability of an experiment to achieve the required accuracy of primordial $B$-mode measurements.

In order to derive calibration requirements for a future CMB satellite mission we developed a simple top-down framework to generate sky maps that take into account the presence of foregrounds, the photometric calibration and band-pass response uncertainties. 
Integrating Equation \ref{eq:bandpassOnly} and writing explicitly each component of our sky model in Equations \ref{eq:dust}, \ref{eq:synch} we find:
\begin{equation}\label{eq:intBandpass}
\begin{split}
    d & = g\Big[I_{cmb}(\nu_0)+\gamma_{d}I_{d}(\nu_0)+\gamma_{s}I_{s}(\nu_0)\Big] + \\ & + g\Big[Q_{cmb}(\nu_0)+\gamma_{d}Q_{d}(\nu_0)+\gamma_{s}Q_{s}(\nu_0)\Big]\cos2\psi + \\ & + g\Big[U_{cmb}(\nu_0)+\gamma_{d}U_{d}(\nu_0)+\gamma_{s}U_{s}(\nu_0)\Big]\sin2\psi + n,
\end{split}
\end{equation}
where the subscript \textit{d} refers to the thermal dust component and the subscript \textit{s} refers to the synchrotron component. Assuming the CMB dipole \cite{dipole1, dipole2} as the natural photometric calibrator for a satellite mission, we define \textit{g} as the photometric calibration factor and $\gamma_{d,s}$ as the color correction factors accounting for the different spectral shape of dust and synchrotron compared to the calibrator (CMB dipole). In Equation \ref{eq:intBandpass}, $\nu_0$ is the effective central frequency of the frequency band.  
If there is no effect other than the band-pass response to take into account, the calibration factor $g$ (using the CMB dipole as calibrator) is determined from the data. Therefore, it does not depend explicitly on the band-pass response knowledge.

If the photometric calibration uncertainty (defined here as $\delta_g$) is negligible, the dominant contribution to take into account is the color correction effect:
\begin{equation}\label{eq:colorCorrection}
    \gamma_{d,s}=\Bigg(\frac{\int d\nu~G(\nu) \frac{I_{d,s}(\nu)}{I_{d,s}(\nu_0)}}{\int d\nu~G(\nu) \frac{\partial B(\nu, T)}{\partial T}\Big|_{T_0}}\Bigg)\frac{\partial B(\nu_0, T)}{\partial T}\Big|_{T_0},\footnote{In both integrals a term $\nu^{-2}$ due to the telescope effective collective area $A_e$ is left implicit, however it is taken into account when performing the calculation.}
\end{equation}
which strongly depends on the prior knowledge of the band-pass response. An incorrect or poor characterization of the band-pass response can cause systematic leakage of foreground signal into the final CMB maps. We define the error on the color correction factor estimation as:
\begin{equation}\label{deltaGamma}
    \delta^{(d,s)}_{\gamma} = \frac{\gamma^{'}_{d,s}-\gamma^{0}_{d,s}}{\gamma^{0}_{d,s}},
\end{equation}
where $\gamma^{0}_{d,s}$ is the color correction factor for an ideal infinite precision for the knowledge of the band-pass response, and $\gamma^{'}_{d,s}$ is the color correction factor for a realistic finite band-pass resolution.

In this section we have shown that photometric calibration and band-pass resolution accuracy are strongly related through Equation \ref{eq:intBandpass}. We can ultimately write the maps for a single detector $j$:

\begin{equation}\label{eq:scanMore}
\begin{pmatrix}
I_j \\
Q_j \\
U_j \\
\end{pmatrix} = 
g\begin{bmatrix}
\begin{pmatrix}
I_{cmb} \\
Q_{cmb} \\
U_{cmb} \\
\end{pmatrix} +
\gamma_d
\begin{pmatrix}
I_{d} \\
Q_{d} \\
U_{d} \\
\end{pmatrix} +
\gamma_s
\begin{pmatrix}
I_{s} \\
Q_{s} \\
U_{s} \\
\end{pmatrix}
\end{bmatrix} + n,
\end{equation}
and use this relation to propagate the calibration uncertainty directly at the map level.


Using this formalism, partially inherited from \cite{Hoang_2017}, we are able to propagate analytically the presence of systematic effects in the maps without computationally expensive time-ordered data simulation for all detectors. 
\subsection{Propagation of the uncertainty at map level} 
\label{sec:calibration}
We can combine the single detector maps in Equation \ref{eq:scanMore} to obtain $I$, $Q$ and $U$ maps for each frequency band, and study the global effect of photometric calibration or band-pass accuracy. If the detectors in the frequency band $i$ are uncorrelated (calibration uncertainties), and the calibration accuracy per detector
(either the photometric calibration factor $g$ or the color correction factor $\gamma$) is known with a precision $\delta_{g,i}$ ($\delta_{\gamma,i}$ for color correction), we can
\begin{figure}[htpb]
	\centering
	\includegraphics[width=1\textwidth]{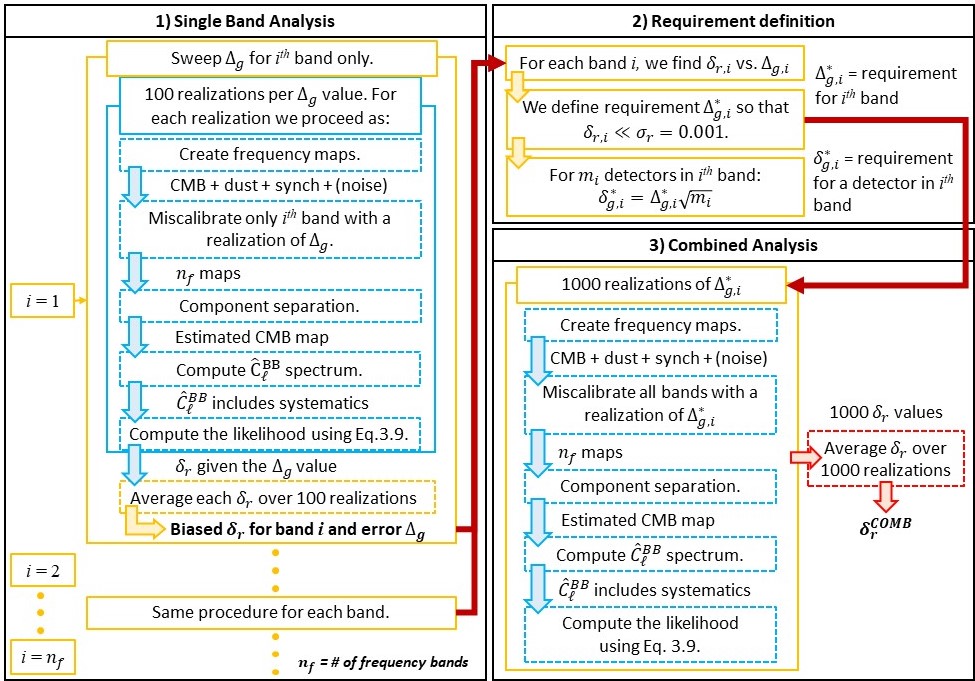}
	\caption{Flow chart of the analysis method followed in this work to study the effect of calibration uncertainty in presence of foreground contamination. The analysis is divided in 3 main steps: single frequency band analysis, requirement determination for each frequency band, and finally requirement validation.}
	\label{analysis}
\end{figure}
propagate the uncertainty to the final (full mission) frequency maps as:
\begin{equation}
    \label{eq:cal_fact}
    \Delta_{[g, \gamma],i}=\frac{\delta_{[g, \gamma],i}}{\sqrt{m_i}}
\end{equation}
where $m_i$ is the number of detectors in the frequency band $i$. Using this factor we can propagate the calibration or color correction error into the maps following Equation \ref{eq:scanMore}, where: 
\begin{equation}
    \label{eq:cal_fact_2}
    [g, \gamma]_i=1+\Delta_{[g, \gamma],i}.
\end{equation}
With this definition we can analytically generate contaminated sky maps by multiplying the ideal sky map at frequency $i$ (including CMB, thermal dust and synchrotron) by the factor $g_i$ ($\gamma_i$ for color correction).
\begin{table}[htpb]
\centering
\begin{tabular}{|c|c|c|c|c|}
    \hline
    \# & $\nu_c$ (GHz) & Bandwidth (Frac.) & \# Detectors & Pol. sensitivity ($\mu K amin$)\\
    \hline
    \hline
    1 & 40 & 12 (30\%) & 64 & 39.76 \\
    \hline
    2 & 50 & 15 (30\%) & 64 & 25.76\\
    \hline
    3 & 60 & 14 (23\%) & 64 & 20.69\\
    \hline
    4 & 68 & 16 (23\%) & 208 & 12.72\\
    \hline
    5 & 78 & 18 (23\%) & 208 & 10.39\\
    \hline
    6 & 89 & 20 (23\%) & 208 & 8.95\\
    \hline
    7 & 100 & 23 (23\%) & 530 & 6.43\\
    \hline
    8 & 119 & 36 (30\%) & 632 & 4.30\\
    \hline
    9 & 140 & 42 (30\%) & 530 & 4.43\\
    \hline
    10 & 166 & 50 (30\%) & 488 & 4.86\\
    \hline
    11 & 195 & 59 (30\%) & 640 & 5.44\\
    \hline
    12 & 235 & 71 (30\%) & 254 & 9.72\\
    \hline
    13 & 280 & 84 (30\%) & 254 & 12.91\\
    \hline
    14 & 337 & 101 (30\%) & 254 & 19.07\\
    \hline
    15 & 402 & 92 (23\%) & 338 & 43.53\\
    \hline
    \end{tabular}
\caption{Configuration for the CMB satellite mission LiteBIRD we assumed in this study, from \cite{LiteBIRD_Sugai}. Central frequency, number of detectors and sensitivity to polarization forecasts are given for all frequency bands. These are the values assumed in the analysis presented in this paper. We need to point out that the final results depend strongly on the instrumental model values. For a different instrumental configuration we shell find different results from those reported in the following.}
\label{tab:sensitivity}
\end{table}

\section{Analysis procedure}
\label{sec:systematics}
In this section we describe the analysis procedure followed in this study. Figure \ref{analysis} shows the flow chart of the analysis procedure.
First, we artificially inject the calibration uncertainty in the frequency maps using the $g_i$ factors, as defined in Section \ref{sec:formalism}, then we use a parametric foreground cleaning method \cite{MaxLikeStompor2, MaxLikeStompor3, likelihood, FgBuster} to study the impact of the calibration uncertainty on the recovered CMB maps. Computing the $B$-mode angular power spectrum we estimate the contamination level in terms of bias on the tensor-to-scalar ratio. We describe the details of the procedure in the following sections.

We explicitly describe the procedure for the photometric calibration analysis. To avoid redundancy we omit the explicit explanation of the color correction analysis, the same procedure can be followed substituting $\Delta_g$ with $\Delta_{\gamma}$.

\subsection{Map preparation}
\label{subsec:sky}
First, we generate full mission sky maps (we make use of the \textit{PySM} library, see \cite{PySM, PysmGit} for details) for all frequency bands in Table \ref{tab:sensitivity} including the sky components mentioned in Section \ref{skyModel}. For the cosmological parameters we adopt the values reported by the Planck experiment in \cite{CosmologicalParam}. Since the goal of LiteBIRD is a target total uncertainty $\sigma_r\leq 0.001$ for $r=0$, we adopt $r=0$ as input to generate the CMB maps.

The sky maps are perturbed as described in Section \ref{sec:calibration}, and specifically making use of Equation \ref{eq:scanMore} to simulate the effect of an imperfect calibration. A white and isotropic noise component is added according to the sensitivity values in Table \ref{tab:sensitivity}. In this paper we assume LiteBIRD baseline parameters as reported in \cite{LiteBIRD_Sugai}. This procedure can be applied to other instrument configurations by changing the instrumental parameters according to a specific design.

Following the formalism introduced in \cite{Ward_2018}, we define for each frequency band $i$ the calibration factor $g_i$ with uncertainty $\Delta_{g,i}$ as:
\begin{equation}
    \label{Gfactors}
    g_{i}=1+\mathcal{N}(0,\Delta_{g,i}),
\end{equation}
where $\mathcal{N}(\mu,\sigma)$ is a random number generated with a Gaussian distribution with mean value $\mu=0$ and standard deviation $\sigma=\Delta_{g,i}$. 

\subsection{Analysis steps}
As mentioned in Section \ref{sec:intro} in \cite{Ward_2018}, a similar analysis is presented for the Simons Observatory case. However, the authors do not make a distinction between the frequency bands. We decided to proceed in the following way to determine which bands are more sensitive to calibration uncertainties, and to reduce the complexity of a 15 free parameters analysis (see also Figure \ref{analysis} for a flow chart of the analysis procedure):
\begin{enumerate}
    \item First, we choose a few representative values for $\Delta_{g,i}$ between $10^{-6}$ and $10^{-2}$. To verify a quadratic relation (see Section \ref{sec:singfreq} for details) between the calibration accuracy of each frequency band ($\Delta_{g,i}$) and the induced bias to the tensor-to-scalar ratio ($\delta_r$), we perform 15 separate analysis (one for each frequency band). For each analysis step, we propagate the uncertainty only for one frequency band (mis-calibration) using the factors from Equation \ref{Gfactors}. After the component separation step, we can relate the uncertainty value directly to the computed $\delta_r$ from the excess in the component separation residuals.
    \item Once we find the relation between $\Delta_{g,i}$ and $\delta_r$ for each frequency band $i$, we define requirements for each band ($\Delta^{*}_{g,i}$) that reduce the bias below the target sensitivity: $\delta_{r} \ll \sigma_r \leq 0.001$. For this work we arbitrarily decided to define the requirement $\delta_{r}\leq5.7\times 10^{-6}$. This value corresponds to $\lesssim1$\% of the target $\sigma_r$ of the experiment. The reason for this choice will be explained in the following. However, the results can be easily re-scaled by the reader for any given requirement.
    \item Lastly, we perform a final simulation propagating the calibration uncertainty in all frequency bands simultaneously using the $\Delta^{*}_{g,i}$ values from the requirement definition at the previous point. Since there are 15 frequency bands and the uncertainties are uncorrelated, we expect to find a total bias to the tensor-to-scalar ratio ($\delta^{COMB}_r$) roughly $\sqrt{15}$ times higher than the threshold value mentioned at the previous point ($\delta_{r}\leq5.7\times 10^{-6}$). However, because of the single frequency requirement $\delta_{r} \leq 5.7\times 10^{-6} \ll \sigma_r$, we expect this to be true for the total tensor-to-scalar ratio bias: $\delta^{COMB}_{r} \ll \sigma_r$.
\end{enumerate}

\subsection{Component separation}
For the component separation part we make use of the \textit{FgBuster} code \cite{FgBuster}, which is a \textit{Python} implementation of the maximum likelihood foreground estimation algorithm described in \cite{likelihood}.

The sky, observed in multiple frequency bands, is modeled at map level, pixel-by-pixel as:
\begin{equation}\label{compsep1}
d_p=A_p s_p+n_p
\end{equation}
where $p$ denotes a single sky pixel, $d_p$ is the observed signal vector (including $n_s=3$ Stokes parameters for $n_f$ frequency bands), $s_p$ is the real sky signal vector ($n_s$ Stokes parameters for $n_c$ number of components), $n_p$ is the noise vector and $A_p\equiv A_p(\beta_i)$ is the mixing matrix of the form $(n_f \cdot n_s)\times (n_c \cdot n_s)$. The mixing matrix is parameterized with the free parameters $\beta_i$ describing the spectrum of each component (see Section \ref{skyModel}). In this analysis we consider 3 components: CMB, synchrotron and thermal dust, therefore we have 3 unknown spatially-uniform parameters $\beta_s$, $\beta_d$ and $T_d$. For $p$ pixels we can remove the subscript and re-write:
\begin{equation}\label{compsep2}
\boldmath d=A s+n.
\end{equation}
Defining the symmetric block diagonal noise matrix $\boldsymbol{N}$ we write the likelihood function as
\begin{equation}\label{compsep3}
-2\ln{\mathcal{L}(\boldsymbol{s},\boldsymbol{\beta})}=\mbox{const}+(\boldsymbol{d}-\boldsymbol{As})^t\boldsymbol{N}^{-1}(\boldsymbol{d}-\boldsymbol{As}),
\end{equation}
the full data likelihood is found as the sum of the likelihood for each single pixel and is maximized when
\begin{equation}\label{compsep4}
\boldsymbol{s}=(\boldsymbol{A}^t\boldsymbol{N}^{-1}\boldsymbol{A})^{-1}\boldsymbol{A}^t\boldsymbol{N}^{-1}\boldsymbol{d},
\end{equation}
then substituting Eq. \ref{compsep4} in Eq. \ref{compsep3} we find:
\begin{equation}\label{compsep5}
-2\ln{\mathcal{L}(\boldsymbol{s},\boldsymbol{\beta})}=\mbox{const}+(\boldsymbol{A}^t\boldsymbol{N}^{-1}\boldsymbol{d})^t(\boldsymbol{A}^t\boldsymbol{N}^{-1}\boldsymbol{A})^{-1}(\boldsymbol{A}^t\boldsymbol{N}^{-1}\boldsymbol{d}).
\end{equation}
The algorithm finds the set of parameters \{$\boldsymbol{\beta_i}$\} that maximize the likelihood function. For more details see \cite{likelihood}. 


\subsection{Tensor-to-scalar ratio bias}
After component separation we obtain an estimated CMB map that is the sum of the input CMB map $m^{true}_{cmb}$, and residuals due to noise $m_{n}$ and the component separation method itself $m_{fg}$. Understanding how to improve the component separation efficiency is the subject of other studies \cite{ErrardStompor2019}. Here we investigate the role that instrumental systematic effects have in boosting the amplitude of the component separation residuals $m_{fg}$.

To determine the residuals due to the calibration uncertainty, we perform a parallel analysis for every value of $\Delta_g$ and noise realization. With the same components and noise maps we run two separate simulations, one with an ideal instrument unaffected by calibration uncertainty ($\Delta_g=0$) and the other propagating the uncertainty $\Delta_g$ into the maps. 

The recovered CMB maps from these two parallel analysis, $m^0_{cmb}$ ($\Delta_g=0$) and $m_{cmb}$ ($\Delta_g\neq 0$) respectively, are differentiated to obtain the residuals map. The residuals map is then analyzed to compute the $B$-mode power spectrum due to the calibration uncertainty (bias spectrum):
\begin{equation}
\Delta m=m_{cmb}-m^0_{cmb} \Rightarrow \hat{C}^{BB}_{\ell,\sigma}.
\label{recovered}
\end{equation}

Finally, to find the bias on the tensor-to-scalar ratio due to the calibration uncertainty, we use the probability distribution function for a measured $B$-mode power spectrum $\hat{C}^{BB}_{\ell}$ and a given value of $r$ \cite{hamimece, Katayama_Komatsu2011}:
\begin{equation}
-2\ln{\mathcal{L}(\hat{C}^{BB}_{\ell}|r)}=\frac{2\ell+1}{2}f_{sky}\Bigg[ \frac{\hat{C}^{BB}_{\ell}}{C^{BB}_{\ell}}+\ln{C^{BB}_{\ell}}-\frac{2\ell-1}{2\ell+1}\ln{\hat{C}^{BB}_{\ell}} \Bigg].
\label{probability}
\end{equation}
The measured signal $\hat{C}^{BB}_{\ell}$ is here equal to the residuals $BB$ power spectrum in Equation \ref{recovered}, while $C^{BB}_{\ell}=rC^{GW}_{\ell}+C^L_{\ell}+N^{BB}_{\ell}+C^{BB}_{\ell,fg}$ is the expected spectrum. Where, $C^{GW}_{\ell}$ is the primordial $B$-mode spectrum (computed for $r=1$), $C^{L}_{\ell}$ is the lensing $B$-mode spectrum, $N^{BB}_{\ell}$ is the noise spectrum and $C^{BB}_{\ell,fg}$ is the residuals spectrum due to foreground components.
We determine the bias to the tensor-to-scalar ratio by finding the peak of:
\begin{equation}
\ln{\mathcal{L}(r)}=\sum^{\ell_{max}}_{\ell=2}\ln{\mathcal{L}(\hat{C}^{BB}_{\ell}|r)},
\label{likelihood}
\end{equation}
where $\ell_{max} = 200$ (assuming this as LiteBIRD $\ell$-range). The tensor-to-scalar ratio bias value is defined as the $r$ value corresponding to the maximum of the likelihood function. 
With this procedure we can express the calibration uncertainty effect in terms of a bias value on the recovered tensor-to-scalar ratio: $\delta_r$ vs. $\Delta_g$ (or $\Delta_{\gamma}$). 

\section{Results}
\label{sec:results}
In this section we present the results of the analysis. We divide the results into 3 subsections, following the steps described in Section \ref{subsec:sky}.
\begin{figure}[htpb]
	\centering
	\begin{minipage}{.5\textwidth}
	\centering
	\includegraphics[width=1\textwidth]{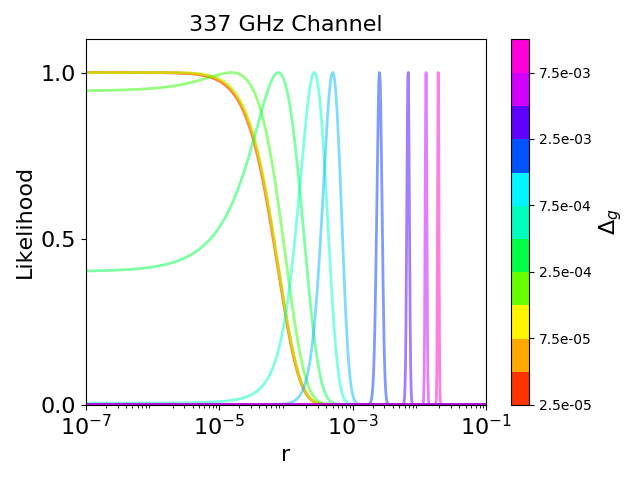}
	\end{minipage}%
	\begin{minipage}{.5\textwidth}
	\centering
	\includegraphics[width=1\textwidth]{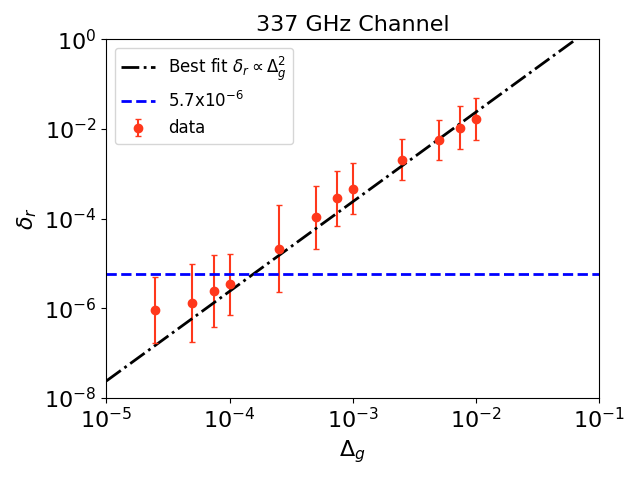}
	\end{minipage}
	\caption{\textit{Left:} Likelihood functions for different values of the $\Delta_g$ factor for one frequency band (337 GHz). \textit{Right:} For each value of $\Delta_g$ we performed 100 simulations with different noise realizations, computed the tensor-to-scalar ratio likelihood for all of them, and calculated the mean $\delta_r$ values. Results are shown for 337 GHz band. We find that the mean $\delta_r$ scales with the square of $\Delta_g$, as expected. The small departure from square-law for small $\Delta_g$ values is due to the finite grid step of the likelihood function calculation.}
	\label{337GHz_gVSr}
\end{figure}

\subsection{Single frequency band analysis}
\label{sec:singfreq}
In this first part of the analysis, we propagate the photometric calibration error one frequency band per time and we study the impact on the data after component separation. In Figure \ref{337GHz_gVSr} we give an example of the analysis. In the \textit{left} panel we show the likelihood functions from the residuals map as explained in Equation \ref{probability}, for different values of $\Delta_g$ for the 337 GHz band.

For each band and for a given value of $\Delta_g$,
we perform 100 simulations varying the noise seed randomly and the effective $\Delta_g$ with a Gaussian distribution with mean value $0$ and standard deviation equal to $\Delta_g$, as explained in Section \ref{subsec:sky}. We define the bias to the tensor-to-scalar ratio ($\delta_r$) due to the calibration error as the $r$ value corresponding to the peak of the likelihood function. For clarity we plot in the \textit{left} panel of Figure \ref{337GHz_gVSr} only one representative curve for each $\Delta_g$ value.

In the \textit{right} panel of Figure \ref{337GHz_gVSr}, for the same frequency band (337 GHz), we computed the mean value of $\delta_r$ over all 100 realizations for each $\Delta_g$ value. We can clearly observe a square-law relation between the computed $\delta_r$ and the input $\Delta_g$ value.
This relation comes from the fact that $\Delta_g$ is a multiplicative factor on the polarization map. Therefore we expect the miscalibration to propagate to the residual polarization map as:
\begin{equation}\label{eq:quadratic1}
    [Q\pm iU](\theta, \phi)\propto\Delta_g\sum_{\ell,m}{}_{\pm2}a_{\ell,m}{}_{\pm2}Y_{\ell,m}(\theta,\phi).
\end{equation}
\begin{figure}[htpb]
	\centering
	\includegraphics[width=.8\textwidth]{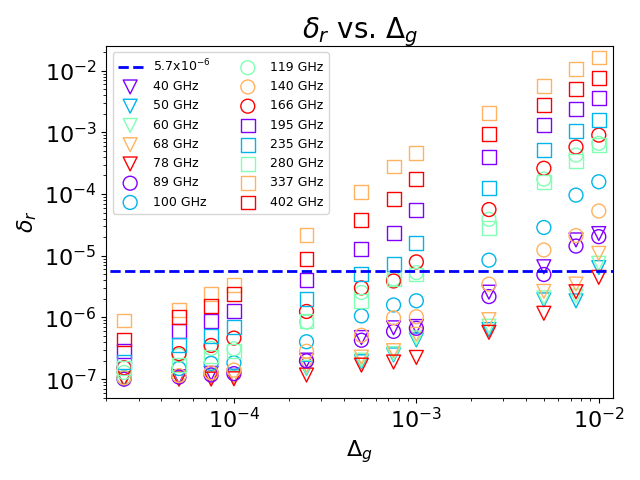}
	\caption{A summary of the first step of the analysis showing the $\delta_r$ vs. $\Delta_g$ relation for every LiteBIRD frequency band.}
	\label{gVSr}
\end{figure}
We find that the spherical harmonic expansion coefficients of the residual $B$-mode map to be:
\begin{equation}\label{eq:quadratic2}
    a_{B,\ell m}\propto\Delta_g\frac{i}{2}({}_{2}a_{\ell,m}-{}_{-2}a_{\ell,m}),
\end{equation}
and thus the angular power spectrum is proportional to $\Delta_g^2$ as:
\begin{equation}\label{eq:quadratic3}
    C_{\ell}^{BB}\propto\sum_{m}a_{B,\ell m}a^{*}_{B,\ell m}\propto\Delta^2_g.
\end{equation}
The relation in Equation \ref{eq:quadratic3} is exact in the limit of a component separation process that does not absorb and compensate calibration (photometric or band-pass) mismatches through the parameters of the model.

In Figure \ref{gVSr}, we show a summary of the results of the analysis for every frequency band in Table \ref{tab:sensitivity}. We can notice that the relation $\delta_r \propto \Delta_g^2$ holds for every frequency band\footnote{The small deviation visible for small $\Delta_g$ values for some of the frequency channels is due to the finite resolution used in the likelihood calculation: $\Delta r=10^{-7}$.}. We also notice that for a given $\Delta_g$ value, the computed $\delta_r$ decreases from low frequency bands towards central frequency bands and then it increases again at higher frequencies. This is due to the relative amplitude of the Galactic foregrounds compared to CMB, which leads to higher residuals in the CMB map. 
We want to draw the attention of the reader to the two highest frequency bands, 337 and 402 GHz, which clearly drive the requirements. As we can see in Figure \ref{gVSr}, the band at 337 GHz appears to be slightly more sensitive to calibration errors compared to the band at 402 GHz. 
This frequency band has higher weight in the component separation procedure, which may be the cause of higher impact on the final $\delta_r$ value.
\begin{table}[htbp]
    \centering
    \begin{tabular}{|| c || c | c |}
    \hline
    Band (GHz) &  $\Delta_{g,\gamma}$  &   $\delta_{g,\gamma}$ \\
    \hline
    \hline
    $40$  \mbox{ } & \mbox{ } $2.5\times 10^{-3}$  \mbox{ } & \mbox{ }   $2.0\times 10^{-2}$  \mbox{ } \\
    \hline
    $50$  \mbox{ } & \mbox{ }  $7.5\times 10^{-3}$  \mbox{ } & \mbox{ }   $6.0\times 10^{-2}$  \mbox{ } \\
    \hline
    $60$  \mbox{ } & \mbox{ }  $7.5\times 10^{-3}$  \mbox{ } & \mbox{ }   $6.0\times 10^{-2}$  \mbox{ } \\
    \hline
    $68$  \mbox{ } & \mbox{ }  $7.5\times 10^{-3}$  \mbox{ } & \mbox{ }   $10.8\times 10^{-2}$  \mbox{ } \\
    \hline
    $78$  \mbox{ } & \mbox{ }  $1.0\times 10^{-2}$  \mbox{ } & \mbox{ }   $14.4\times 10^{-2}$  \mbox{ } \\
    \hline
    $89$  \mbox{ } & \mbox{ }  $5.0\times 10^{-3}$  \mbox{ } & \mbox{ }   $7.2\times 10^{-2}$  \mbox{ } \\
    \hline
    $100$  \mbox{ } & \mbox{ }  $1.0\times 10^{-3}$  \mbox{ } & \mbox{ }   $2.3\times 10^{-2}$  \mbox{ } \\
    \hline
    $119$  \mbox{ } & \mbox{ }  $1.0\times 10^{-3}$  \mbox{ } & \mbox{ }   $2.5\times 10^{-2}$  \mbox{ } \\
    \hline
    $140$ \mbox{ } & \mbox{ }  $2.5\times 10^{-3}$  \mbox{ } & \mbox{ }   $5.7\times 10^{-2}$  \mbox{ } \\
    \hline
    $166$ \mbox{ } & \mbox{ }  $7.5\times 10^{-4}$  \mbox{ } & \mbox{ }   $1.6\times 10^{-2}$  \mbox{ } \\
    \hline
    $195$ \mbox{ } & \mbox{ }  $2.5\times 10^{-4}$  \mbox{ } & \mbox{ }   $0.6\times 10^{-2}$  \mbox{ } \\
    \hline
    $235$ \mbox{ } & \mbox{ } $5.0\times 10^{-4}$  \mbox{ } & \mbox{ }   $0.8\times 10^{-2}$  \mbox{ } \\
    \hline
    $280$ \mbox{ } & \mbox{ } $1.0\times 10^{-3}$  \mbox{ } & \mbox{ }   $1.6\times 10^{-2}$  \mbox{ } \\
    \hline
    $337$ \mbox{ } & \mbox{ }  $1.0\times 10^{-4}$  \mbox{ } & \mbox{ }   $0.16\times 10^{-2}$  \mbox{ } \\
    \hline
    $402$ \mbox{ } & \mbox{ } $1.0\times 10^{-4}$  \mbox{ } & \mbox{ }   $0.18\times 10^{-2}$  \mbox{ } \\
    \hline
    \end{tabular}
    \caption{A summary of the requirements in terms of the overall frequency bands ($\Delta_{g,\gamma}$), and per detector ($\delta_{g,\gamma}$) assuming the number of detectors in Table \ref{tab:sensitivity}.}
    \label{tab:req}
\end{table}

\subsection{Requirements}
\label{subsec:req}
The goal of LiteBIRD, as for most next generation CMB experiments, is to measure the tensor-to-scalar ratio with high accuracy. In this work we assume $r=0$, and we assume a total target uncertainty $\sigma_r \leq 0.001$. To do so, we have to reduce the cumulative bias of all the systematic effects to a negligible value. 

Since we expect to have multiple systematic effects in an experiment, we arbitrarily decide a threshold value $\delta_r\lesssim 5.7\times 10^{-6}$ which is negligible compared to the target sensitivity.\footnote{We assume that the total uncertainty can be divided in $\sigma_{fg}$ due to the component separation residuals and $\sigma_{sys}$ due to systematic effects. By requiring that these two terms have the same value, including a margin term $\sigma_{m}$ and combining them in quadrature to obtain the total uncertainty, we define the uncertainty value allocated to each term: $\sigma_{i}\sim5.7\times 10^{-4}$. Since we expect the experiment to suffer from multiple systematic effects, we assign to each effect 1\% of the total systematic budget.} This value is marked with a blue dashed line in both Figure \ref{337GHz_gVSr} and Figure \ref{gVSr}. Following this procedure we can define a requirement $\Delta^{*}_{g,i}$ for each frequency band that satisfies the condition $\delta_r(\Delta^{*}_{g,i})\lesssim 5.7\times 10^{-6}$. A summary of the requirements is given in Table \ref{tab:req}. From the number of detectors per band in Table \ref{tab:sensitivity} we also derive the requirement per detector $\delta^{*}_{g,i}$ using Equation \ref{eq:cal_fact}. It is important to remind the reader that the detector requirements in Table \ref{tab:req} are 
valid if the detector calibration uncertainties are uncorrelated.

\subsection{Combined analysis}
Finally we propagate the calibration errors in all 15 bands at the same time, using the $\Delta^{*}_{g,i}$ values of Table \ref{tab:req}. For a perfectly linear procedure and uncorrelated errors we expect the cumulative mean bias to be equal to $\sqrt{15}\times 5.7\times 10^{-6}$. 

\begin{figure}[htpb]
	\centering
	\includegraphics[width=.6\textwidth]{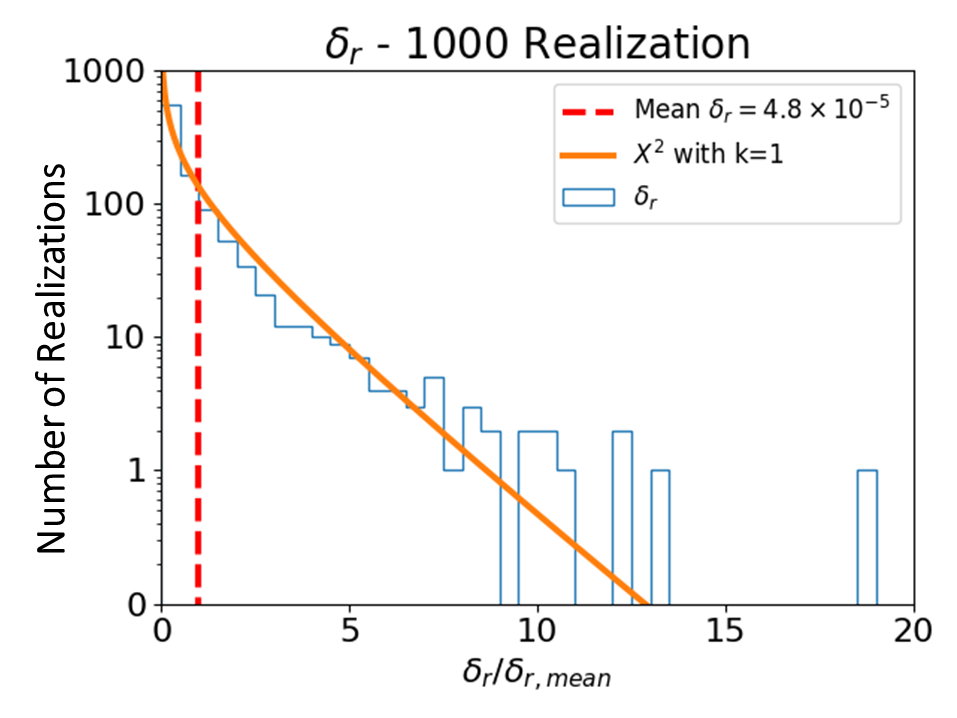}
	\caption{Summary of the 1000 simulations where we propagate the calibration error in all frequency bands at the same time using the values in Table \ref{tab:req}. We compute the bias to the tensor-to-scalar ratio (as the $r$ value corresponding to the peak of the likelihood function) for each of the 1000 simulations, and we plot the distribution in units of the mean value $\delta_{r,mean}$. The mean value (corresponding to $1$ in these units) is shown as a vertical dashed \textit{red} line. A $\chi^2$-distribution with $k=1$ degree of freedom is shown for comparison (the residuals amplitude). }
	\label{rBiasAll}
\end{figure}

For this final step of the analysis, we perform 1000 simulations varying the random noise seed and the $\Delta^{*}_{g,i}$ realizations. A summary of the results can be seen in Figure \ref{rBiasAll}. As expected, the total tensor-to-scalar ratio bias $\delta^{COMB}_r=4.8\times10^{-5}$ (mean value over 1000 realizations 
in Figure \ref{rBiasAll}) is higher than the threshold value defined in Section \ref{subsec:req}, but slightly in excess than the expected value for a perfect linear system of equations. However, the component separation step is a non-linear process. 

\begin{figure}[htpb]
	\centering
	\begin{minipage}{.5\textwidth}
	\centering
	\includegraphics[width=0.95\textwidth]{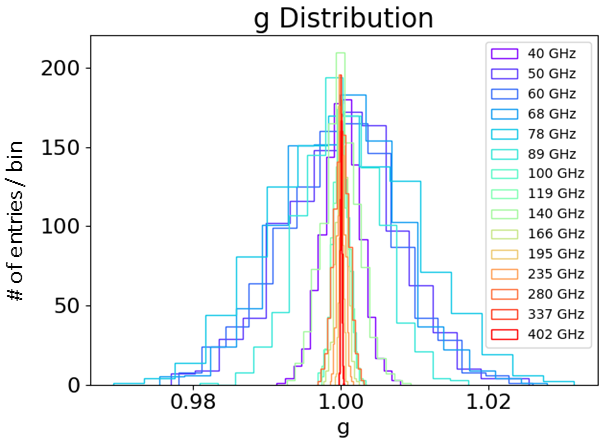}
	\end{minipage}%
	\begin{minipage}{.5\textwidth}
	\centering
	\includegraphics[width=0.95\textwidth]{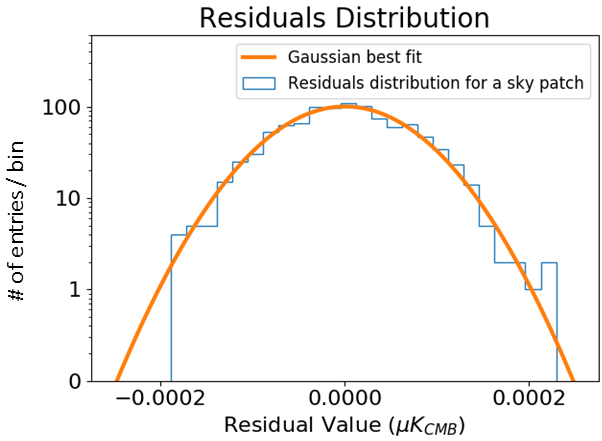}
	\end{minipage}
	\caption{\textit{Left:} Distribution of the $g$-factors used in the final steps of the analysis presented in this paper. \textit{Right:} Distribution of the mean Q residual value (similar distribution for U) over 1000 simulations in a sky patch. We perform a Gaussian fit of the data.}
	\label{GandResDistrib}
\end{figure}

In Figure \ref{rBiasAll}, we show the distribution of the computed tensor-to-scalar ratio values for all 1000 simulations. The $g$-factors are randomly drawn from a normal distributions with standard deviations $\Delta^{*}_{g,i}$. Thus, if the component separation process was perfectly linear, we would expect the amplitude of the residual maps to follow a normal distribution as well. Figure \ref{GandResDistrib} shows the distributions of the 1000 $g$-factors used in the simulation and the distribution of the amplitude of the residuals in the CMB maps. Through this statement and Equations \ref{eq:quadratic1}, \ref{eq:quadratic2} and \ref{eq:quadratic3}, we find that the $\delta_r$ distribution has to follow a $\chi^2$-distribution with 1 degree of freedom (the residuals amplitude). In Figure \ref{rBiasAll} we can see a fairly good agreement between the data and the $\chi^2$-distribution, apart for a small excess in the tail which induces the excess in the computed mean value $\delta^{COMB}_r$.

Table \ref{tab:req} reports the calibration requirements for each frequency band. 
As explained in the previous sections, these requirements are found by imposing a threshold value for the bias $\delta_r\leq 5.7\times 10^{-6}$, assuming at each step that only one frequency band suffers from calibration errors.

We carried out the same analysis for the color correction factor $\gamma_{[s,d]}$. We found the same requirements as for the photometric calibration factor $g$ in Table \ref{tab:req}, and thus we avoid reporting them separately.

\section{Discussion}\label{sec:discussion}
Through this analysis, we find that for the LiteBIRD case 2 are the most sensitive bands to calibration errors (337 and 402 GHz), therefore we can take these values as the overall experiment requirement. In this way we can reduce the contribution from the other bands to a negligible level and achieve a total bias $\delta^{COMB}_r \sim 5.7\times 10^{-6}$.

A second point worth mentioning is that by increasing the number of detectors at high frequency it is possible to reduce the single detector requirement, although this solution might be impractical due to cost issues, especially in the case of a space mission. 

\subsection{Band-pass}
Lastly, we use these results to address the color correction effect and the implication on the required band-pass response resolution.

In Equation \ref{eq:colorCorrection} we defined the calibration correction factor (color correction) for the Galactic foreground signals, and in Equation \ref{deltaGamma} we defined the error for the color correction factor for a finite band-pass resolution.
For a given finite resolution $\Delta\nu$ of the band-pass response $G(\nu)$, Equation \ref{eq:colorCorrection} becomes:
\begin{equation}\label{eq:gammaFinite}
    \gamma_{d,s}=\Bigg(\frac{\sum_i \Delta\nu G(\nu_i) \frac{I_{d,s}(\nu_i)}{I_{d,s}(\nu_0)}}{\sum_i \Delta\nu G(\nu_i) \frac{\partial B(\nu_i, T)}{\partial T}\Big|_{T_0}}\Bigg)\frac{\partial B(\nu_0, T)}{\partial T}\Big|_{T_0}.
\end{equation}
As already mentioned, similarly to what we have presented in Section \ref{sec:results} for the calibration factor $g$, we performed the analysis for the color correction factors $\gamma$. We avoid reporting the full analysis here because the requirements found are consistent to those reported for $g$ in Section \ref{sec:results}. This correspondence between the two results comes from the fact that the combined Galactic foreground components dominate the CMB $B$-mode signal at all frequencies. Therefore, if the cosmological signal is negligible, the effect of $g$ and $\gamma$ is indistinguishable in Equation \ref{eq:scanMore}. Using the $\delta_g$ value reported in Table \ref{tab:req}, we can find the required band-pass resolution to minimize the color correction error and the bias effect on the recovered tensor-to-scalar ratio. 

If we knew the band-pass response of our instrument with infinite precision, the color correction error, as defined in Equation \ref{deltaGamma}, would be zero. Under this circumstances, we would be able to perfectly calibrate the foreground signals with respect to CMB. Therefore we define requirements that minimize the error and the impact on the data.
To illustrate our procedure, we focus on the band most sensitive to calibration error as found in the previous sections: 337 GHz. At this frequency the dominant sky component is thermal dust, therefore we limit our analysis to this foreground component for the remaining of this section. The required calibration accuracy for a single detector in this frequency band is $0.16\times 10^{-2}$ (see Table \ref{tab:req}), therefore we need to find the band-pass resolution ($\Delta\nu$) that satisfies this requirement.
\begin{figure}[htpb]
	\centering
	\begin{minipage}{.475\textwidth}
	\centering
	\includegraphics[width=1\textwidth]{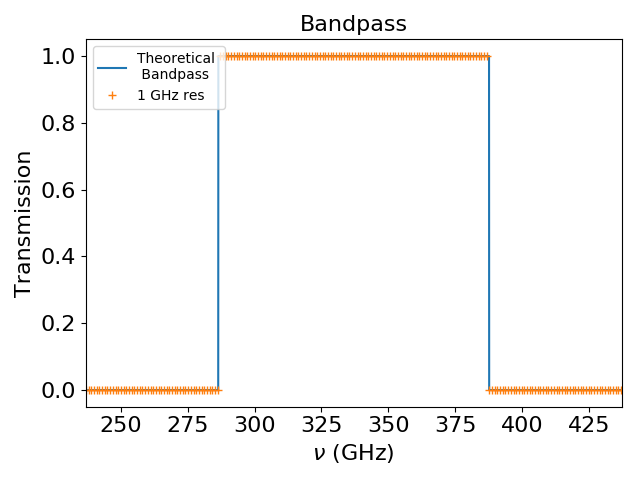}
	\end{minipage}%
	\begin{minipage}{.475\textwidth}
	\centering
	\includegraphics[width=1\textwidth]{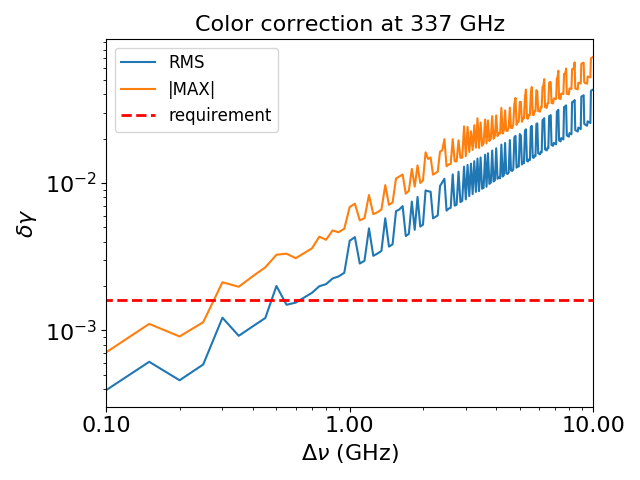}
	\end{minipage}
	\begin{minipage}{.475\textwidth}
	\centering
	\includegraphics[width=1\textwidth]{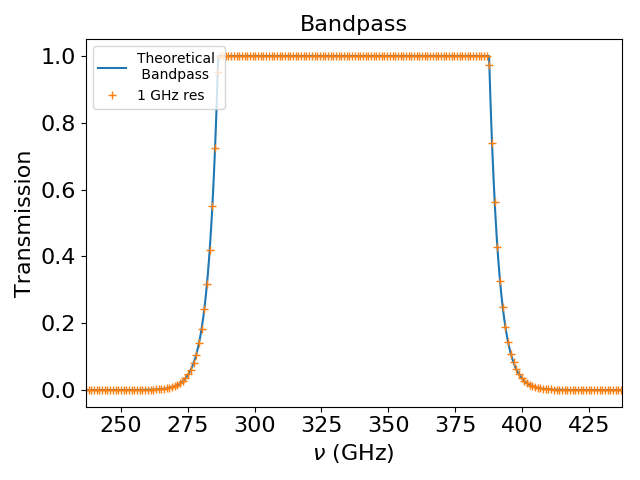}
	\end{minipage}%
	\begin{minipage}{.475\textwidth}
	\centering
	\includegraphics[width=1\textwidth]{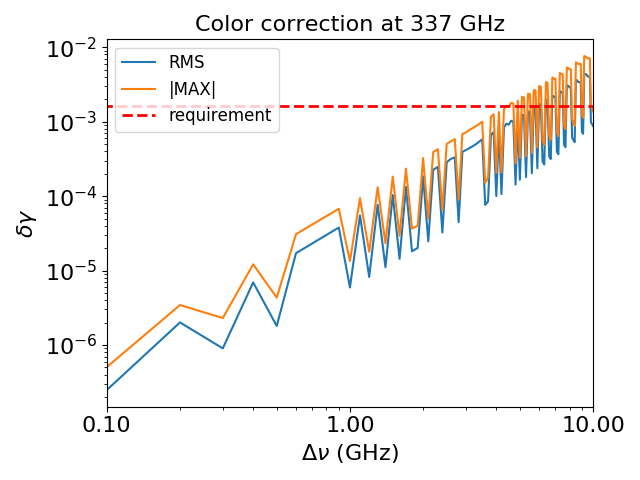}
	\end{minipage}
	\begin{minipage}{.475\textwidth}
	\centering
	\includegraphics[width=1\textwidth]{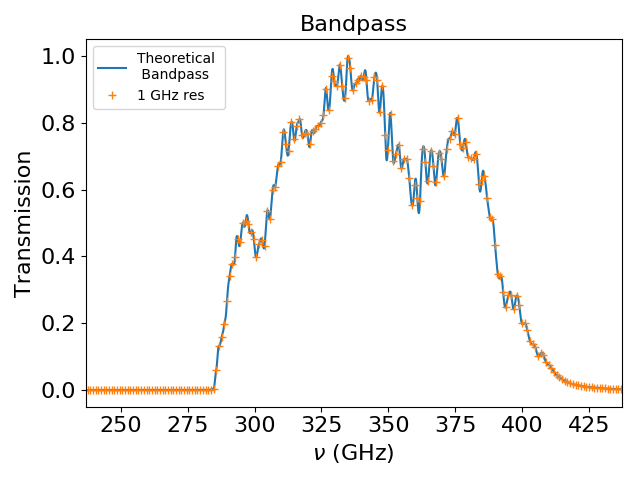}
	\end{minipage}%
	\begin{minipage}{.475\textwidth}
	\centering
	\includegraphics[width=1\textwidth]{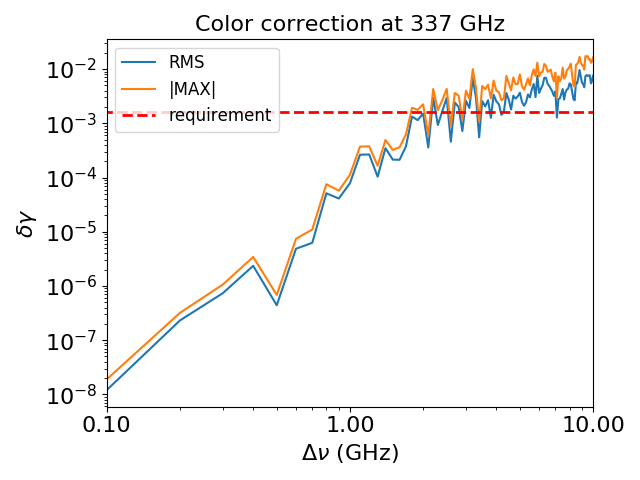}
	\end{minipage}
	\caption{\textit{Left:} In \textit{blue} solid line examples of three theoretical band-pass responses at 337 GHz with 30\% bandwidth are shown. From \textit{top} to \textit{bottom}, we show a perfect top-hat ideal band-pass response, a top-hat with a more realistic transitions at the edges, and finally, using publicly available Planck data \cite{planckSupplement, planckArchive} we re-scaled one of the Planck 353 GHz band-pass responses to 337 GHz. A re-sampling process with 1 GHz resolution is also shown as a scattered plot. \textit{Right:} Calculation of the color correction error for dust ($\delta^{d}_{\gamma}$) as a function of decreasing resolution for the band-pass response on the left. The \textit{blue} solid line represents the \textit{rms} value for 100 realizations of the re-sampling process with a given resolution, while the \textit{orange} solid line represents the maximum value between 100 realizations. The requirement shown by the \textit{red} dashed line.}
	\label{fig:bandpassGamma}
\end{figure}

In Figure \ref{fig:bandpassGamma}, we analyze three cases to show how the shape of the band-pass response influences the calculation of $\delta_{\gamma}$. For all cases we create a theoretical band-pass response with almost infinite resolution ($\Delta\nu = 0.1$ MHz)\footnote{Current and past experiments have reported measuring the band-pass response with $\sim1$ GHz resolution \cite{matsuda2019polarbear}, and therefore a 0.1 MHz resolution is a good approximation for a nearly infinite resolution.}. Then, we proceed with re-sampling the band-pass response with lower resolution and we compute $\delta_{\gamma}$ for the new resolution through Equation \ref{deltaGamma}. In Figure \ref{fig:bandpassGamma} \textit{top-left}, \textit{center-left} and \textit{bottom-left} we plot the three representative theoretical band-pass responses in \textit{blue} solid line and a re-sampling step of the same function with 1 GHz resolution as an example.
\begin{figure}[htpb]
	\centering
	\begin{minipage}{.5\textwidth}
	\centering
	\includegraphics[width=1\textwidth]{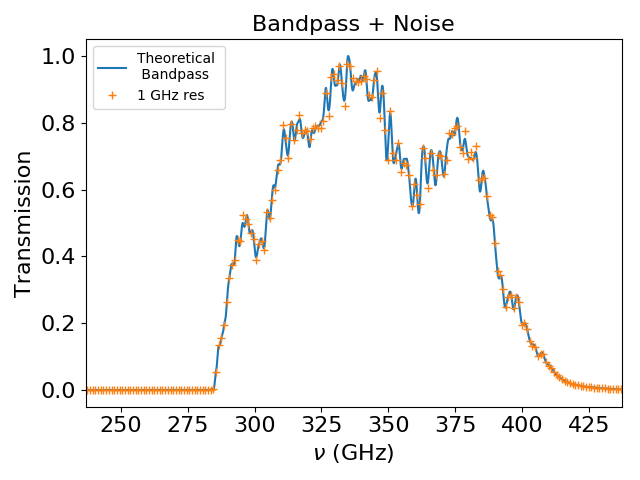}
	\end{minipage}%
	\begin{minipage}{.5\textwidth}
	\centering
	\includegraphics[width=1\textwidth]{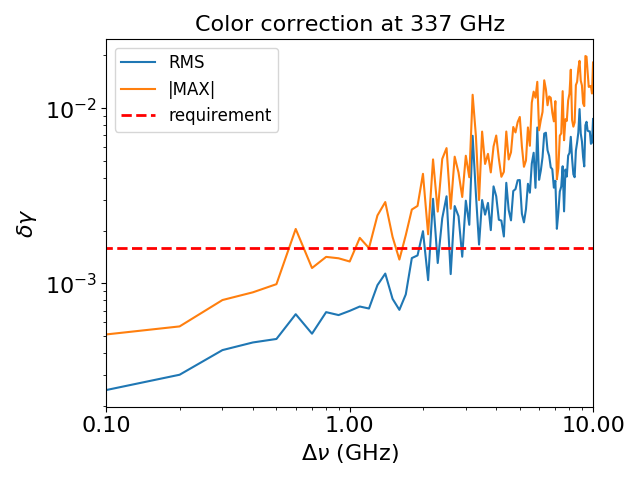}
	\end{minipage}
	\caption{\textit{Left:} In \textit{blue} solid line a theoretical band-pass responses at 337 GHz with 30\% bandwidth, simulated using publicly available Planck data \cite{planckSupplement, planckArchive}, is shown. We re-scaled one of the Planck 353 GHz band-pass responses to 337 GHz. A re-sampling process with 1 GHz resolution is also shown as a scattered plot. A white noise measurement component is simulated during the re-sampling process \textit{Right:} Calculation of the color correction error for dust ($\delta^{d}_{\gamma}$) as a function of decreasing resolution for the band-pass response. The \textit{blue} solid line represents the \textit{rms} value for 100 realizations of the re-sampling process with a given resolution, while the \textit{orange} solid line represents the maximum value between 100 realizations. The requirement is shown by the \textit{red} dashed line. The statistical uncertainty boosts $\delta_{\gamma}$ especially for high resolution.}
	\label{fig:bandpassGammaNoise}
\end{figure}

The \textit{top-left} panel of Figure \ref{fig:bandpassGamma} shows a top-hat band-pass, which presents a perfectly flat in-band response with sharp transitions between the in-band region and the out-of-band regions. Although this ideal band shape is not realistic, in the \textit{top-right} panel of Figure \ref{fig:bandpassGamma} we show that the sharp transitions at the edges of the band-pass response impact negatively the color correction factor. In this case a very fine resolution is required to reduce the uncertainty. 

A more realistic case is shown in Figure \ref{fig:bandpassGamma} \textit{center-left}. This case still presents a flat response in-band, while the edge transitions are smoother to mimic a more realistic case. This choice is completely arbitrary and the calculation has been done purely to show the effect of smoother transitions compared to a top-hat response. The \textit{center-right} panel of Figure \ref{fig:bandpassGamma} clearly shows that the edge smoothness helps reducing the uncertainty of the $\gamma$ factor. 

Finally, we create a more realistic band-pass response using publicly available Planck data at 353 GHz \cite{planckSupplement, planckArchive}. We shift and re-scale the band-pass response to match the 337 GHz central frequency, as shown in the \textit{bottom-left} panel of Figure \ref{fig:bandpassGamma}. Figure \ref{fig:bandpassGamma} \textit{bottom-right} shows $\delta_{\gamma}$ as a function of the resolution. The original resolution of the Planck data is $\sim 2$ GHz, and therefore we have to interpolate the data to simulate a higher resolution. Because of this choice there is a lack of information in the Planck filter for resolution higher than $2$ GHz. This effect can be seen in Figure \ref{fig:bandpassGamma} \textit{bottom}, where the $\delta_{\gamma}$ as a function of the sampling resolution becomes much steeper below $\sim 2$ GHz. From these results we find that a resolution 0.2 GHz $\lesssim\Delta\nu\lesssim$ 2 GHz is necessary to achieve an error lower than the requirement threshold, depending on the effective shape of the band-pass function. 

This procedure can be followed for every other frequency band and for other Galactic foreground emission. We avoid reporting all results here because the one just shown for thermal dust at 337 GHz is the most stringent.


\subsection{Some considerations}
\paragraph{Band-pass resolution.} The calculation color correction depends strongly on the effective shape of the band-pass response. The usual zero-order approximation for a band-pass response is a top-hat function centered at the nominal central frequency. This shape increases dramatically the error of the color correction factor because of the steep transitions between in-band and out-of-band regions, while a shallower transition helps reducing the error. Another source of error are fast fringes in the response, caused by standing waves between optical elements of the telescope. Given these considerations we recommend future experiments to carefully and realistically simulate the the spectral response of the system to fully understand the impact that this might have on the observations. 

We also need to point out the presence of statistical uncertainty and systematic effects within the set-up to measure the band-pass response. The first, depending on the statistical noise level, will limit the high resolution end of Figure \ref{fig:bandpassGamma} \textit{right}, and therefore increasing the resolution will not help in reducing the color correction uncertainty as shown in Figure \ref{fig:bandpassGammaNoise}. On the other end, systematic effect in the set-up can create artificial features in the measured response that will result in an incorrect measure of the band-pass response. As an example in Figure \ref{fig:bandpassGammaNoise} \textit{left}, we report a case similar to \ref{fig:bandpassGamma} \textit{bottom-left}, where we artificially add a 2\% white noise component to the band-pass data. This value is in line with the results reported by the POLARBEAR collaboration in \cite{matsuda2019polarbear}\footnote{In \cite{matsuda2019polarbear} a signal-to-noise ratio $S/N\sim20$ has been reported, which would correspond to $\sim5\%$ noise level. However, the authors break down the uncertainty into a statistical component and systematic component, and they identify the former to be $\sim2\%$ in the worst case reported. Since we consider here only a statistical component we decided to use the $\sim2\%$ value. From Equation \ref{eq:gammaFinite} we can easily find that the $\delta_{\gamma}$ scales linearly with an uncertainty on $G$, therefore the requirement can be quickly re-scaled by the reader for a different noise level.}. As it is visible by comparing Figures \ref{fig:bandpassGamma} \textit{bottom-right} and \ref{fig:bandpassGammaNoise} \textit{right}, the color correction factor error deteriorates in presence of statistical uncertainty especially for very fine resolution. 

\paragraph{Calibration.} Throughout this paper, we assumed that a space mission like LiteBIRD will make use of the CMB dipole signal as the primary photometric calibrator for every frequency band. 
The dipole is the most accurate known photometric calibrator for a CMB space-borne mission because it is well characterized, and therefore, it is possible to achieve high calibration accuracy. Other possible calibrators, like planets for example (or an artificial calibrator, see \cite{CalSat}) present higher uncertainties and therefore the accuracy might suffer from such a choice. In addition, using non-extended sources (like planets) calibration will become more sensitive to beam uncertainties (far-side lobes, etc.).

In the event that a different calibrator is going to be used for some of the analysis, Equation \ref{eq:intBandpass} has to be adapted to the new calibrator, because this new calibrator will have most certainly a different spectrum compared to the CMB and the dipole signals (planets are "dusty" sources, so probably a grey-body spectrum has to be assumed); therefore a different color correction scheme needs to be taken into account.

\paragraph{Sky model.} Throughout this paper, we assumed a fairly simple sky model with only two Galactic components, without spatial variation of the parameters or synchrotron curvature. A more complex sky might induce higher residuals due to the complexity of the sky, making more complicated the separation of the residuals due to the component separation method itself, and those due to the systematic effect under study \cite{ErrardStompor2019}. On the other end, a more complex sky model (e.g. with multiple dust populations) with more parameters to be fitted, might have the effect of absorbing more efficiently the calibration errors, resulting in a relaxation of the requirements. 

\paragraph{Component separation method.} In this paper we applied a parametric method to perform the component separation procedure, however other existing methods could be applied. An analysis of the difference between the various methods is beyond the scope of this paper, but should definitely be explored in the future by the community (a comparison of the methods available at the time of the Planck mission can be found in \cite{Leach_2008}). A potential advantage of ILC (Internal Linear Combination) methods may be to relax the high frequency requirements driven by the dust emission, given the fact that these methods are relying solely on assumptions about the cosmological signal \cite{Bennett_2003, Delabrouille2009}. We can foresee that miscalibration affecting only the foreground components (like the color correction) may be re-absorbed without propagating to the estimation of the CMB signal and the tensor-to-scalar ratio. However, as shown in \cite{Dick_2010}, ILC methods tend to suffer if the CMB signal is miscalibrated. Further investigation of the strengths and weaknesses of various methods is needed in the future.

\paragraph{Uncorrelated noise.} In this analysis we assumed uncorrelated uncertainties among detectors. For what concerns the assumption of uncorrelated detector noise this is justified by the fact that if we consider correlations between detectors the sensitivity per band assumed in Table \ref{tab:sensitivity} needs to be re-scaled according to the level of correlation. Therefore, this analysis needs to be repeated for the new sensitivity which will impact the efficiency of the component separation procedure. On the other hand, techniques have been adopted in the past to mitigate the presence of correlated noise \cite{Patanchon_2008, deGasperis_2016}. Secondly, even in the presence of correlated noise among detectors we do not have evidence, at present, of the possible presence of correlations in the bandpass uncertainty which justify the use of Equation \ref{eq:cal_fact} to re-scale the uncertainty. This should be reconsidered if and when we will have evidence of the contrary.

\paragraph{1/f noise.} 1/f noise is certainly going to be one of the challenges for LiteBIRD or any other next generation CMB surveys. However, assuming a HWP rotating at $\sim 1$ Hz shifts the polarized cosmological signal at higher frequency $\sim 4$ Hz, where the noise level is likely to be uncontaminated by the 1/f component. Details of the 1/f mitigation using a continuously rotating HWP can be found in \cite{KusakaABS1, Takakura_2017}. LiteBIRD will make use of a rotating HWP to mitigate the 1/f component, therefore in this analysis we considered the polarized signal to be unaffected by the possible presence of 1/f noise. However, in future the origin and magnitude of the 1/f component and its possible impact on the data needs to be carefully studied.

\section{Conclusions}
\label{sec:conclusion}

In conclusion, instrument calibration is a fundamental step in defining the complete instrument model necessary for a correct data analysis in CMB experiments. In this paper we have shown how to determine the impact of photometric calibration uncertainty for a space-borne CMB polarimeter using an ideal polarization modulator (like LiteBIRD). The use of a polarization modulator is justified by the requirement of mitigating and controlling systematic effects and reducing the impact of $1/f$ noise. However, even such a system will not be immune to systematic effects that need to be understood. In this paper we have specifically focused on the effect of photometric calibration uncertainty and imperfect knowledge of the band-pass response even when a rotating HWP is employed as a polarization modulator. 

We have presented a method to address the impact of photometric calibration uncertainty in the presence of Galactic foregrounds. We have derived requirements for the calibration accuracy to minimize the effect on the data. In particular we discussed the effect of a finite band-pass response resolution. Starting from plausible instrument parameters taken from the baseline design of the LiteBIRD satellite, we simulated the effect of an imperfect calibration, and showed that for the set of parameters in Table \ref{tab:sensitivity}, the high frequency bands (specifically 337 and 402 GHz) are the most sensitive to calibration uncertainty ($\delta_g$). We found requirements per frequency band: $\Delta_{g,\gamma}\sim10^{-4}-2.5\times10^{-3}$, and per single detector: $\delta_{g,\gamma}\sim0.18\times10^{-2}-2.0\times10^{-2}$. We found a quadratic relation between the calibration uncertainty $\Delta_g$ and the tensor-to-scalar ratio bias $\delta_r$. To illustrate our method we defined a $\delta_r$ threshold to identify a requirement on $\Delta_g$, which can be easily re-scaled for a different choice of the threshold. We also modelled the effect of a finite band-pass resolution, and derived requirements for it in order to minimize the effect of a limited band-pass knowledge. We adopted a few representative examples for the band-pass response and computed the resolution requirement: $\Delta\nu\sim 0.2 - 2$ GHz (depending on the band-pass shape assumed). Although the Planck-like band-pass is more representative of a real scenario than the other two cases analysed, the derived requirement of $\Delta\nu\sim2$ GHz might suffer from the limited resolution of the original data. On the other end, given the unrealistic sharpness of the top-hat case, we can fairly assume the 0.2 GHz resolution requirement, for the 337 GHz band, as the worst case scenario.  
Ultimately, these results highlight the need for carefully modelling the response of the telescope system to help the definition of the calibration requirements. 

\section{Acknowledgment}
TG thanks University of Oxford and Kavli IPMU for the financial support throughout his doctoral studies. TG, TM and MH acknowledge the World Premier International Research Center Initiative (WPI), MEXT, Japan for support through Kavli IPMU. This work was supported by JSPS KAKENHI Grant Numbers JP15H05891 and 18KK0083, as well as by the JSPS Core-to-Core Program, A. Advanced Research Networks. We thank Josquin Errard and Davide Poletti for useful discussion about the parametric component separation method, for giving us access to the {\itshape FgBuster} code and for useful comments on the manuscript. In this work we made extensive use of the HEALPIx package \cite{Gorski_2005}. We also thank the KEK Computing Research Center for granting access to KEKCC facility.

\newpage

\bibliographystyle{JHEP}
\bibliography{bibi}



\end{document}